\documentclass[11pt]{article}

\usepackage{amsmath}
\usepackage{graphicx}
\usepackage{amssymb}
\usepackage{amsthm}

\newtheorem{theorem}{Theorem}[section]
\newtheorem{lemma}[theorem]{Lemma}
\newtheorem{corollary}[theorem]{Corollary}
\newtheorem{proposition}[theorem]{Proposition}
\numberwithin{equation}{section}

\begin{document}

\title{Uniqueness theorems for equations of Keldysh Type}
\author{Thomas H. Otway\thanks{email: otway@yu.edu} \\
\\
\textit{Department of Mathematics, Yeshiva University}\\ \textit{New York,
NY 10033 USA}}
\date{}
\maketitle

\begin{abstract}
A fundamental result that characterizes elliptic-hyperbolic
equations of Tricomi type, the uniqueness of classical solutions to
the open Dirichlet problem, is extended to a large class of
elliptic-hyperbolic equations of Keldysh type. The result implies
the non-existence of classical solutions to the closed Dirichlet
problem for this class of equations. A uniqueness theorem is also
proven for a mixed Dirichlet-Neumann problem. A generalized
uniqueness theorem for the adjoint operator leads to the existence
of distribution solutions to the closed Dirichlet problem in a
special case. \textit{MSC2000}: 35M10
\end{abstract}

\section{Introduction}

In 1956, Morawetz \cite{Mo1} proved the uniqueness of smooth
solutions to  \emph{open} Dirichlet problems, having data prescribed
on only part of the boundary, for certain mixed elliptic-hyperbolic equations
of Tricomi type. That result implied that the \emph{closed}
Dirichlet problem, in which data are prescribed on the entire
boundary, is over-determined for smooth solutions of such equations.

Morawetz's result was later extended to a large class of boundary
value problems for Tricomi-type equations, by Manwell (\cite{Ma1}
and Sec.\ 16 of \cite{Ma2}) and by Morawetz herself \cite{Mo3}. But there
is as yet no corresponding result for equations of Keldysh type --
the other canonical local form for second-order linear equations of
mixed elliptic-hyperbolic type. Morawetz's result has been extended
to a particular equation which is of Keldysh type in any
neighborhood of the origin and of Tricomi type elsewhere \cite{MSW}.
That proof requires the type-change function to be symmetric about
the $x$-axis and an analytic function of its arguments, neither of
which we assume. In Sec.\ 2 we generalize the assertion of
\cite{MSW} to an entire class of equations under mild restrictions
on the type-change function. While that result is by no means
unexpected, its apparent absence from the literature until now,
nearly 80 years after the derivation by Cinquini-Cibrario
of the two canonical forms for elliptic-hyperbolic equations \cite{CC}, is
rather surprising.

Morawetz considered equations essentially having the form
\begin{equation}\label{trity}
    \mathcal{K}(y)u_{xx}+u_{yy}=0,
\end{equation}
where
\begin{equation} \label{cond1}
\mathcal{K}(0)=0
\end{equation}
and
\begin{equation} \label{tricond2}
y\mathcal{K}(y)>0 \mbox{ for } y\ne 0.
\end{equation}
Here and below, the unknown function $u$ depends on
$\left(x,y\right)\in\mathbb{R}^2.$ Equations having the form (\ref{trity}), possibly including lower-order terms and satisfying conditions (\ref{cond1}) and (\ref{tricond2}), are said to be of
\emph{Tricomi type}. Equations having the form
\begin{equation}\label{keldysh}
    \mathcal{K}(x)u_{xx}+u_{yy}+ \mbox{ lower-order terms }=0,
\end{equation}
where $\mathcal{K}(x)$ satisfies (\ref{cond1}) and
\begin{equation}\label{cond2}
    x\mathcal{K}(x)>0 \mbox{ for } x\ne 0,
\end{equation}
are said to be of \emph{Keldysh type}.

In Sec.\ 2 we consider a special case of these equations, namely,
equations having the form
\begin{equation}\label{loword}
Lu \equiv \mathcal{K}(x)u_{xx}+u_{yy}+\frac{\mathcal{K}'(x)}{2}u_x=0,
\end{equation}
where $\mathcal{K}$ satisfies conditions (\ref{cond1}) and
(\ref{cond2}). We assume for convenience that $\mathcal{K}$ is
$C^1,$ and monotonic on the hyperbolic region, but the result
clearly extends to weaker hypotheses on $\mathcal{K}.$ For example, the
monotonicity hypothesis on $\mathcal{K}$ is imposed only in order to
simplify the graphs of the characteristic lines, which in turn simplifies the proof of Theorem \ref{theorem0} in the hyperbolic region. See also the discussion of eq.\ (\ref{lavbit}), below.

As an example, choose $\mathcal{K}(x) = x^{2k_0-1}$ for $k_0\in
\mathbb{Z}^+.$ The operator $L$ under this choice of type-change
function is roughly analogous, for equations of Keldysh type, to the
well known \emph{Gellerstadt operator} \cite{Ge} for equations of
Tricomi type. Other examples are the polar-coordinate forms of the
equation for harmonic fields on the extended projected disc (eq.\
(16) of \cite{O}) and an equation arising from a uniform asymptotic
approximation of high-frequency waves near a caustic (eq.\ (4.1) of
\cite{MT}). Both equations can be put into the form (\ref{loword})
in the cartesian $r\theta$-plane, with the $x$-axis replaced by the
line $r=1.$

The main object of Sec.\ 2 is to show that any smooth solution of
eq.\ (\ref{loword}) which vanishes identically on the
non-characteristic boundary of a typical domain also vanishes in the
interior (Theorem \ref{theorem0}). This implies the nonexistence of
classical solutions to a closed Dirichlet problem (Corollary
\ref{corollary1}). We also show that an analogous result holds if
only the normal derivative of the solution vanishes on the
horizontal arcs of the non-characteristic boundary (Theorem
\ref{mixed}).

In the important special cases in which $\mathcal{K}(x)=x$ or $L$ is formally self-adjoint, we prove
uniqueness theorems (in a very generalized sense) for solutions to
the adjoint equation. This allows us to show the existence of
distribution solutions to a closed Dirichlet problem by so-called
\emph{projection} methods (Theorems \ref{theorem1} and
\ref{theorem2}). Given these results, it is natural to wonder about
the existence of weak solutions to closed boundary value problems. This is a difficult question. Some
obstacles to proving the existence of weak solutions to a closed
Dirichlet problem by the direct application of even powerful
projection methods are discussed in Sec.\ 5.1 of \cite{O1}. However, the existence of weak solutions to the closed Dirichlet problem for the equation studied in \cite{MSW} has been proven \cite{O3}.

\section{The nonexistence of classical solutions}

For given $\mathcal{K}(x),$ define constants $a,$ $b,$ $d,$ and $m,$ where $m<a \leq 0<d$ and $b>0.$ Consider the domain $\mathcal{D}$ formed by
the line segments
\[\
\mathcal{L}_1 = \left\{\left(x,y\right)|
a\leq x\leq d, y=-b\right\};
\]
\[
\mathcal{L}_2 =
\left\{\left(x,y\right)|x=d, -b\leq y\leq b\right\};
\]
\[
\mathcal{L}_3 =
\left\{\left(x,y\right)|a\leq x\leq d, y=b\right\};
\]
the characteristic line $\Gamma_1$ joining the points
$\left(m,0\right)$ and $\left(a,-b\right);$ and the characteristic
line $\Gamma_2$ joining the points $\left(m,0\right)$ and
$\left(a,b\right).$

The $y$-axis divides the domain $\mathcal{D}$ of eq.\
(\ref{loword}) into the subdomains
\[
\mathcal{D}^+=\left\{\left(x,y\right)\in\mathcal{D}|x\geq0\right\},
\]
and $\mathcal{D}^-=\mathcal{D}\backslash\mathcal{D}^+.$ Equation
(\ref{loword}) is (non-uniformly) elliptic for
$\left(x,y\right)\in\mathcal{D}^+.$

\begin{theorem} \label{theorem0} Let $u\left(x,y\right)$ be a
twice-differentiable solution of eq.\ (\ref{loword}), with
$\mathcal{K}$ satisfying conditions
(\ref{cond1}) and (\ref{cond2}). Assume that $\mathcal{K}$ is $C^1,$ and monotonic on $\mathcal{D}^-.$ If $u$
vanishes on the non-characteristic boundary, then $u\equiv 0$ on all of $\mathcal{D}.$
\end{theorem}

\begin{proof} We follow the approach of \cite{Mo1}, \cite{Mo3}, and \cite{MSW}. Introducing the auxiliary function
\[
I =
\int_0^{\left(x,y\right)}\left[\mathcal{K}(x)u_x^2-u_y^2\right]dy-2u_xu_ydx,
\]
we compute
\[
\frac{\partial}{\partial
y}\left(-2u_xu_y\right)-\frac{\partial}{\partial
x}\left[\mathcal{K}(x)u_x^2-u_y^2\right]
\]
\[
=\left(-2u_{xy}u_y-2u_xu_{yy}\right)-\left[\mathcal{K}'(x)u_x^2+2\mathcal{K}(x)u_xu_{xx}-2u_yu_{yx}\right]
\]
\[
=-2u_xu_{yy}-\left[\mathcal{K}'(x)u_x^2+2\mathcal{K}(x)u_xu_{xx}\right]
\]
\[
=-2u_x\left(u_{yy}+\frac{\mathcal{K}'(x)}{2}u_x+\mathcal{K}(x)u_{xx}\right)
=0,
\]
using (\ref{loword}) and the equivalence of mixed partial
derivatives. We conclude that there exists a function $\xi\left(x,y\right)$ such
that
\begin{equation} \label{xix}
\xi_x=-2u_xu_y
\end{equation}
and
\begin{equation} \label{xiy}
\xi_y = \mathcal{K}(x)u_x^2-u_y^2.
\end{equation}

We will first show that $u$ vanishes identically in $\mathcal{D^+}.$
To accomplish this, we must show that $u$ vanishes identically on
the sonic line $x=0.$ Once we have shown that, we will have zero boundary
conditions on $\mathcal{D}^+.$ We will complete the proof for the
elliptic region by invoking a maximum principle for non-uniformly elliptic equations.

Because $u\equiv 0$ on $\mathcal{L}_1,$ we conclude that $u_x$
vanishes identically on that horizontal line. Thus we have, by
(\ref{xix}),
\begin{equation} \label{Nu1}
\xi_x = 0 \mbox{ on } \mathcal{L}_1.
\end{equation}
Also, $u\equiv 0$ on $\mathcal{L}_3,$ so $u_x = 0$ on that
horizontal line as well, implying that
\begin{equation} \label{Nu2}
\xi_x = 0 \mbox{ on } \mathcal{L}_3.
\end{equation}
Equations (\ref{Nu1}) and (\ref{Nu2}) imply that, on $\mathcal{L}_1$
and $\mathcal{L}_3,$ $\xi$ is a function of $y$ only. But $y$ is
constant on those two horizontal lines, implying that
\[
\xi = c_1 \mbox{ on } \mathcal{L}_1
\]
and
\[
\xi=c_2 \mbox{ on } \mathcal{L}_3,
\]
where $c_1$ and $c_2$ are constants. On the line $\mathcal{L}_2,$
$u\equiv 0,$ implying that $u_y=0$ on that vertical line. Also,
$\mathcal{K}(x)>0$ on $\mathcal{L}_2.$ These facts imply, using eq.\
(\ref{xiy}), that $\xi_y\geq 0$ on $\mathcal{L}_2,$ which in turn
implies that
\begin{equation} \label{c1}
c_2\geq c_1.
\end{equation}
On the line $x=0,$ $\mathcal{K}=0,$ implying by (\ref{xiy}) that
$\xi_y\leq 0$ on that vertical line. This is turn implies that
\begin{equation} \label{c2}
c_2\leq c_1.
\end{equation}
Inequalities (\ref{c1}) and (\ref{c2}) are in contradiction unless
$c_1=c_2.$ Taking into account that $\xi$ cannot increase with increasing $y$ on
the line $x=0,$ it also cannot decrease with increasing $y,$ as it would then
have to increase in order to return to its initial value at the
endpoint. This implies that $\xi_y=0$ on the $y$-axis. Using
(\ref{xiy}) again, we find that on the $y$-axis,
\begin{equation} \label{ucon}
-u_y^2=0,
\end{equation}
so the function $u\left(0,y\right)$ is constant there. Because
\[
u\left(0,-b\right)=u\left(0,b\right)=0,
\]
that constant is zero. Thus on the rectangle $\partial\mathcal{D}^+$
we have a closed Dirichlet problem having homogeneous boundary
conditions.

A well known extension of the maximum principle to non-uniformly
elliptic operators (Proposition \ref{max}) implies that the smooth function $u$ attains both
its maximum and minimum values on the boundary. Because it is
identically zero there, $u$ must be zero in all of $\mathcal{D}^+.$

We obtain the identical vanishing of $u$ on the hyperbolic region by
integration along characteristic lines as in \cite{ANP}. We have
\[
d\xi=\xi_xdx+\xi_ydy=\left(-2u_xu_y\right)dx+\left[\mathcal{K}(x)u_x^2-u_y^2\right]dy.
\]
On characteristic lines,
\[
    dx=\pm\sqrt{-\mathcal{K}(x)}dy
\]
and
\[
d\xi = \left[\mp 2u_xu_y\sqrt{-\mathcal{K}(x)}+\mathcal{K}(x)u_x^2-u_y^2\right]dy
\]
\begin{equation}\label{chara}
=-\left[\sqrt{-\mathcal{K}(x)}u_x\pm u_y\right]^2dy\leq 0.
\end{equation}
Thus $\xi$ is non-increasing in $y$ on any arbitrarily chosen
characteristic.

Initially, take $a=0.$

Because $u\equiv 0$ on the sonic line $x=0,$ we conclude that $u_y=0$ on that vertical line. So $\xi_x=0$ on the sonic line by (\ref{xix}). Because $K(0)=0,$ we conclude that $\xi_y=0$ on the sonic line by (\ref{xiy}) and (\ref{ucon}). Beginning at the point $\left(0,-b\right),$ proceed along
$\Gamma_1$ to $\left(m,0\right)$ and then along $\Gamma_2$ to
$\left(0,b\right).$ Expression (\ref{chara}) implies that $\xi$
will not increase in $y$ along this path from $\left(0,-b\right)$ and $\left(0,
b\right).$ Because $\xi$ is equal to the same constant at those two
points, $\xi$ must be constant in $y$ along $\Gamma_1\cup\Gamma_2.$
(If $\xi$ decreased in $y$ at any point along such its path, it
would have to increase in $y$ at a later point in order to return to
its constant value at $\left(0,b\right).$ And it cannot increase in
$y$ along a characteristic.) Ascending along the $y$-axis from the
point $\left(0,-b\right),$ for any initial point above $\left(0,-b\right)$ and any terminal point below $\left(0,b\right)$ on the $y$-axis we can always find a pair of
characteristic lines intersecting at some point on the $x$-axis to
the right of $\left(m,0\right).$ We conclude that $\xi_y=0$ on
$\mathcal{D}^-.$ But then (\ref{xiy}) implies that
\begin{equation} \label{equal}
\mathcal{K}(x)u_x^2=u_y^2 \mbox{ on } \mathcal{D}^-.
\end{equation}
Because $\mathcal{K}<0$ on $\mathcal{D}^-,$ we are forced to
conclude from (\ref{equal}) that $u_x=u_y=0$ on $\mathcal{D}^-.$
This is turn implies that $u$ is constant on $\mathcal{D}^-.$
Because $u\equiv 0$ on the sonic line, that constant must be zero by
the smoothness of $u.$

Now take $a<0.$ Because $\xi_x=0$ on $\mathcal{L}_1$ and
$\mathcal{L}_3,$ $\xi$ remains constant between $\left(0,-b\right)$
and $\left(a,-b\right)$ and between $\left(0,b\right)$ and
$\left(a,b\right).$ Moreover, $\xi_y$ remains non-positive along
$\Gamma_1$ and $\Gamma_2.$ As we move the initial and terminal
points to the right along $\mathcal{L}_1$ and $\mathcal{L}_3$ in
$\mathcal{D}^-,$ we can always find a pair of characteristic lines
which intersect at a point on the $x$-axis to the right of
$\left(m,0\right).$ Arguing as in the case $a=0,$ we again conclude
that $u\equiv 0$ on $\mathcal{D}^-.$ This completes the proof of
Theorem \ref{theorem0}.
\end{proof}

\bigskip

In the special case in which $\mathcal{K}$ is an analytic function,
we do not require a maximum principle, so we do not need to show
that $u=0$ on the line $x=0.$ Rather, we observe that $u_y=0$  on
$\mathcal{L}_2$ as $u\equiv 0$ on that vertical line. Our analysis
of the constants $c_1$ and $c_2$ implies that $\xi_y = 0$ on
$\mathcal{L}_2$ as well. Because in addition, $\mathcal{K}>0$ on
$\mathcal{L}_2,$ eq.\ (\ref{xiy}) implies that $u_x = 0$ on
$\mathcal{L}_2.$ We use this last identity as Cauchy data for the
Cauchy-Kowalevsky Theorem, to argue that $u$ remains equal to zero
as one moves in the negative $x$-direction away from $\mathcal{L}_2$
along the rectangle $\mathcal{D}^+.$ This argument was applied in
\cite{MSW}.

An example of a natural type-change function which is \emph{not} analytic is the function
\begin{equation} \label{lavbit}
\mathcal{K}(x) = \mbox{sgn}[x],
\end{equation}
which yields an analogue, for equations of Keldysh type, of the
Lavrent'ev-Bitsadze equation \cite{LB}. Although such
$\mathcal{K}(x)$ is also not $C^1,$ our proof will work for this
choice of $\mathcal{K}$ provided (\ref{loword}) is suitably
interpreted.

\begin{corollary}\label{corollary1}The closed Dirichlet problem for
eq.\ (\ref{loword}) on $\mathcal{D}$ cannot have a
twice-continuously differentiable solution.
\end{corollary}

\begin{proof} Suppose that $u_1$ and $u_2$ are two smooth
solutions of the open Dirichlet problem for (\ref{loword}) on
$\mathcal{D},$ with data prescribed only on $\mathcal{L}_1,$
$\mathcal{L}_2,$ and $\mathcal{L}_3.$ Then $U\equiv u_2-u_1$
satisfies the hypotheses of Theorem \ref{theorem0}. We conclude that
$u_1=u_2$ in $\mathcal{D}.$ That is, any smooth solution to eq.\
(\ref{loword}) is uniquely determined by data given on the
non-characteristic boundary. So the problem is over-determined for
smooth solutions if data are given on the entire boundary. This
completes the proof.
\end{proof}

\begin{theorem} \label{mixed} The conclusion of Theorem \ref{theorem0}
remains true if the Dirichlet conditions on the non-characteristic
boundary of $\mathcal{D}$ are replaced by the following mixed
Dirichlet-Neumann conditions: $u_y\equiv 0$ on $\mathcal{L}_1$
and $\mathcal{L}_3;$ $u\equiv 0$ on $\mathcal{L}_2.$
\end{theorem}

\begin{proof} The existence of $\xi$ satisfying eqs.\ (\ref{xix}) and (\ref{xiy})
is established by the same arguments as in the proof of Theorem
\ref{theorem0}. The condition that $u_y = 0$ on $\mathcal{L}_1$ and
$\mathcal{L}_3$ implies that $\xi_x=0$ on those horizontal lines. So
the proof of Theorem \ref{theorem0} implies that $\xi$ is equal to a
constant $c_0$ on $\mathcal{L}_1$ and $\mathcal{L}_3.$ Because $u=0$
on $\mathcal{L}_2,$ eqs.\ (\ref{xix}) and (\ref{xiy}) imply that
$\xi$ is equal to $c_0$ on $\mathcal{L}_2$ as well. The arguments leading to eq.\
(\ref{ucon}) imply that $u$ is constant on the line $x=0$ (but not
necessarily equal to zero, as we no longer assume the vanishing of
$u$ on the lines $\mathcal{L}_1$ and $\mathcal{L}_3).$ So eqs.\
(\ref{xix}) and (\ref{xiy}) imply that $\xi$ is constant on the line
$x=0.$ Because of the conditions on $\mathcal{L}_1$ and
$\mathcal{L}_3,$ that constant is equal to $c_0.$ Thus we conclude
that $\xi$ is equal to $c_0$ on the rectangle $\partial\mathcal{D}^+.$

A direct calculation, using (\ref{loword}), (\ref{xix}), (\ref{xiy}), and the identity of mixed partial derivatives, shows that $\xi$ satisfies
\[
\mathcal{K}(x)\xi_{xx}+\xi_{yy}+\frac{\mathcal{K}'(x)}{2}\xi_x=0.
\]
Now Proposition \ref{max} implies that $\xi$ is a
constant (not necessarily zero) in $\mathcal{D}^+.$ In particular,
(\ref{xix}) implies that
\[
\xi_x=-2u_xu_y=0,
\]
so $u_x=0$ and/or $u_y=0.$ If $u_x=0,$ then $u\equiv 0$ in $\mathcal{D}^+$ because $u=0$ on $\mathcal{L}_2.$  If $u_y=0,$ then (\ref{xiy}) and the constancy of $\xi$ imply that
\[
\xi_y=\mathcal{K}u_x^2=0.
\]
Because $\mathcal{K}>0$ on
$\mathcal{D}^+\backslash\left\{x=0\right\},$ we conclude that
$u_x=0$ on $\mathcal{D}^+\backslash\left\{x=0\right\}.$ Because
$u=0$ on $\mathcal{L}_2,$ we again conclude that $u\equiv 0$ on
$\mathcal{D}^+\backslash\left\{x=0\right\}.$ The smoothness of $u$
implies that $u$ is also zero on the line $x=0.$

The proof that $u\equiv 0$ in $\mathcal{D}^-$ is the same as in the proof of Theorem \ref{theorem0}.
\end{proof}

\begin{corollary} Let $f_1,$ $f_2,$ and $f_3$ be given functions defined on the arcs $\mathcal{L}_1,$
 $\mathcal{L}_2,$ and $\mathcal{L}_3,$ respectively. The mixed Dirchlet-Neumann problem
 in which $u_y=f_1$ on $\mathcal{L}_1,$ $u_y = f_3$ on $\mathcal{L}_3,$ $u=f_2$ on $\mathcal{L}_2,$ and any
 boundary conditions at all are imposed on the characteristic lines, is over-determined for smooth solutions
 of (\ref{loword}) in $\mathcal{D}.$
\end{corollary}

\section{The existence of distribution solutions}

Consider operators having the form
\begin{equation} \label{kappaop}
L_\kappa = xu_{xx}+\kappa u_x+u_{yy},
\end{equation}
where $\kappa$ is a constant in the interval $\left[0,3/2\right].$
This class includes the operator $L_0$ originally studied by
Cinquini-Cibrario (\cite{CC3}; see also \cite{SXC}). Our first results will apply to a smaller class of operators in which $\kappa$ is at least 1. The formal
adjoint of $L_\kappa$ is given by
\[
L_\kappa^\ast u = xu_{xx}+\left(2-\kappa\right)u_x+u_{yy}.
\]

\begin{lemma} \label{lemma} Denote by $\Omega$ any bounded,
connected subdomain of $\mathbb{R}^2$ having piecewise smooth
boundary with counter-clockwise orientation. Let $u$ be any $C^2$
function on $\Omega$ which vanishes on the boundary $\partial\Omega$
and also satisfies
\begin{equation} \label{L2_minus_alt}
x u_x^2+u_y^2=0
\end{equation}
on $\partial \Omega.$ Then if $\kappa\in\left[1,3/2\right],$
\begin{equation}\label{funda}
    \left[\int\int_\Omega
\left(|x|u_x^2+u_y^2\right)\,dxdy\right]^{1/2}\leq C||L_{\kappa}^\ast
    u||_{L^2(\Omega)}.
\end{equation}
\end{lemma}

\begin{proof} The proof is similar to the proof of Theorem 2 of
\cite{O1}, and is based on ideas in Sec.\ 2 of \cite{LMP}.

For a positive constant $\delta<<1,$ define the function
\[
Mu = au + bu_x + cu_y,
\]
where $a=-1,$ $c=2\left(2\delta-1\right)y,$ and
\[
b= b_1(x)+b_2(y)=\left\{
        \begin{array}{cr}
    \exp\left(2\delta x/Q_1\right) + \left(2\delta-1\right)\left(1-\kappa\right)y^2 & \mbox{if $x\in\Omega^+$} \\

    \exp\left(3\delta x/Q_2\right)+ \left(2\delta-1\right)\left(1-\kappa\right)y^2 & \mbox{if $x\in\Omega^-$}\\
    \end{array}
    \right..
\]
Here $\Omega^+=\left\{x\in\Omega\,|\,x
\geq 0\right\}$ and $\Omega^-=\Omega\backslash\Omega^+.$ Choose
$Q_1=\exp\left(2\delta \mu_1\right),$ where
$\mu_1=\max_{x\in\overline{\Omega^+}}x.$ Define the negative number
$\mu_2$ by $ \mu_2=\min_{x\in\overline{\Omega^-}}x$ and let
$Q_2=\exp\left(\mu_2\right).$ For example, if $\Omega =
\mathcal{D},$ where $\mathcal{D}$ is the domain of Sec.\ 2, then
$\mu_1=d$ and $\mu_2=m.$ (The constants $a$ and $b$ defined in this section have nothing to do with the constants $a$ and $b$ defined in the preceding section.)

Notice that on $\Omega^+,$
\[
2\delta x \leq 2\delta\mu_1\leq
2\delta\mu_1e^{2\delta\mu_1}=Q_1\log Q_1,
\]
or
\[
\frac{2\delta x}{Q_1}\leq\log Q_1.
\]
Exponentiating both sides, we conclude that $b_1\leq Q_1$ on
$\Omega^+.$

Choose $\delta=\delta\left(\Omega\right)$ to be sufficiently small
so that $3\delta<Q_2.$ Then on $\Omega^-,$
\[
3\delta x\geq 3\delta\mu_2=3\delta\log Q_2 > Q_2\log Q_2,
\]
so $b_1 > Q_2$ on $\Omega^-.$

The coefficient $b(x,y)$ exceeds zero for $\kappa\geq 1$ and is continuous but not differentiable on the
$y$-axis. When we integrate over $\Omega,$ it is necessary to
introduce a cut along the $y$-axis, which is analogous to the procedure employed in \cite{LMP}. The boundary
integrals involving $a,$ $b,$ and $c$ on either side of this line
will cancel. Integrating by parts using Proposition \ref{IBP} with $\mathcal{K}(x)=x$ and $k=2-\kappa,$ we obtain
\begin{eqnarray}
\left(Mu,Lu\right)=\frac{1}{2}\oint_{\partial\Omega} \left(x
u_x^2+u_y^2\right)\left(cdx-bdy\right)\nonumber\\
+\int\int_{\Omega^+\cup\Omega^-}\alpha
u_x^2+2\beta u_xu_y+\gamma u_y^2\,dxdy, \label{IBPform}
\end{eqnarray}
where
\[
\alpha_{\Omega^+} = \delta\left[2-\frac{b_1}{Q_1}\right]x
+\left(\frac{3}{2}-\kappa\right)b\geq \delta x;
\]
\[
\alpha_{\Omega^-} =
\delta\left[2-3\frac{b_1}{Q_2}\right]x+\left(\frac{3}{2}-\kappa\right)b
 \geq \delta\vert x\vert;
\]
\[
2\beta = c\left(1-\kappa\right)-b_y = 0;
\]
if $\delta$ is sufficiently small, then there is a positive constant
$\varepsilon$ such that
\[
\gamma_{\Omega^+} = 2+\delta\left(\frac{b_1}{Q_1}-2\right)\geq\varepsilon
\]
and
\[
\gamma_{\Omega^-} =2+\delta\left(\frac{3b_1}{2Q_2}-2\right)\geq\varepsilon.
\]
The path integral in (\ref{IBPform}) vanishes by identity (\ref{L2_minus_alt}).

Let
\[
\delta' = \min\left\{\delta,\varepsilon\right\}.
\]
Then
\[
    \delta'\int\int_\Omega
\left(|x|u_x^2+u_y^2\right)\,dxdy \leq
    \left(Mu,Lu\right)\leq
\]
\[
    ||Mu||_{L^2}||Lu||_{L^2}\leq
    C(\Omega)\left[\int\int_\Omega
\left(|x|u_x^2+u_y^2\right)\,dxdy\right]^{1/2}||Lu||_{L^2(\Omega)},
\]
where we have used Proposition \ref{Poinc} in obtaining the
bound on the $L^2$-norm of $Mu.$ (In the proof of that proposition it is sufficient for $u$ to be $C^1$ and to vanish on $\partial\Omega.$) Dividing through by the
weighted double integral on the right completes the proof of Lemma
\ref{lemma}.
\end{proof}

Obviously, defining a constant $k = 2-\kappa,$ we can replace
inequality (\ref{funda}) with an analogous inequality in which the $L^2$-norm of $L_{\kappa}^\ast
    u$ is replaced by the $L^2$-norm of $L_ku$ for $k\in \left[1/2,1\right].$ In
particular, taking $\kappa=3/2,$ we obtain the inequality
\begin{equation} \label{fun1}
\left[\int\int_\Omega
\left(|x|u_x^2+u_y^2\right)\,dxdy\right]^{1/2}\leq C||L_{1/2}u||_{L^2(\Omega)}
\end{equation}
This yields a proof of Theorem \ref{theorem0} for
$\mathcal{K}(x)=x$ which does not require a maximum principle:

Apply the proof of Lemma \ref{lemma}, taking $\Omega = \mathcal{D}$ and
$\kappa=3/2.$ This choice converts inequality (\ref{funda}) into inequality (\ref{fun1}).  The arguments leading to eq.\ (\ref{ucon}) imply that
condition (\ref{L2_minus_alt}) is satisfied on any boundary arc on
which $u_y=0$ and $\xi,$ as defined by eqs.\ (\ref{xix}) and
(\ref{xiy}), satisfies $\xi_y=0,$ and that those conditions are
satisfied on the boundary arc $\mathcal{L}_2.$ The path integral in (\ref{IBPform})
is nonnegative on the line segments $\mathcal{L}_1,$ $\mathcal{L}_3,$ and the line $x=0$ by the definitions of $b$ and $c$ and the orientation of $\partial\mathcal{D}^+.$ Uniqueness on the elliptic part
of the domain follows from inequality (\ref{fun1}) without applying any
maximum principle.

\medskip

More generally, we can apply Lemma \ref{lemma} to show the existence
of distribution solutions to the equation $L_\kappa = 0.$ (A brief discussion
of the weighted function spaces that we will apply in the remainder of this section
is given in Sec.\ A.2.)

Consider, still more generally, equations having the form
\begin{equation}\label{f-alt1}
    Lu=f,
\end{equation}
where $f$ is a given, sufficiently smooth function of $(x,y).$ By a
\emph{distribution solution} of eq.\ (\ref{f-alt1}) with the
boundary condition
\begin{equation}\label{boundary}
    u(x,y)=0\,\forall (x,y)\in\partial\Omega
\end{equation}
we mean a function $u\in L^2(\Omega)$ such that $\forall \xi \in
H^1_0(\Omega;\mathcal{K})$ for which $L^\ast\xi\in L^2(\Omega),$ we
have
\begin{equation}\label{ds}
    \left(u,L^\ast\xi\right)=\langle f,\xi \rangle.
\end{equation}
Here $\langle \,,\, \rangle$ is the
\emph{duality bracket} (or \emph{duality pairing}); this can be defined from the $H^{-1}$ norm via the formula
\[
||w||_{H^{-1}(\Omega;\mathcal{K})}=\sup_{0\neq\xi\in
C^\infty_0(\Omega)}\frac{|\langle w,\xi
\rangle|}{||\xi||_{H^1_0(\Omega;\mathcal{K})}},
\]
and is motivated by the Schwarz inequality
\[
|\langle w,\xi \rangle| \leq ||w||_{H^{-1}\left(\Omega;\mathcal{K}\right)}||\xi||_{H^1_0\left(\Omega;\mathcal{K}\right)}
\]
for $w\in H^{-1}\left(\Omega;\mathcal{K}\right)$ and $\xi\in H^1_0\left(\Omega;\mathcal{K}\right).$

Notice that distribution solutions to a homogeneous Dirichlet
problem need not vanish on the boundary! Thus we will not use the
argument of Lemma \ref{lemma} to establish uniqueness in the
conventional sense. In particular, we do not need to show directly that
condition (\ref{L2_minus_alt}) is satisfied if $u$ has compact support in $\Omega.$ Rather, we note that
the proof of Lemma \ref{lemma} will also prove the following:

\begin{lemma} \label{lemma_alt} Denote by $\Omega$ any bounded,
connected subdomain of $\mathbb{R}^2$ having piecewise smooth
boundary with counter-clockwise orientation. Let $u$ be any $C^2_0$
function on $\Omega.$ Then inequality (\ref{funda}) is satisfied, and can be written in the form
\begin{equation} \label{fundalt}
||u||_{H^1_0\left(\Omega;x\right)}
\leq C||L_{\kappa}^\ast
    u||_{L^2(\Omega)}
\end{equation}
for $\kappa\in\left[1,3/2\right].$
\end{lemma}

This leads to the following existence result:

\begin{theorem} \label{theorem1} The Dirichlet
problem $L_\kappa u=f$ with boundary condition (\ref{boundary})
possesses a distribution solution $u\in L^2(\Omega)$ for every $f
\in H^{-1}(\Omega;x)$ whenever $\kappa\in\left[1,3/2\right].$
\end{theorem}

\begin{proof} The proof for our case is essentially identical to the proof in the well known case of Tricomi-type operators (\emph{c.f.} \cite{LMP}, Theorem 2.2). Define for
$\xi \in C_0^\infty$ a linear functional
\begin{equation} \label{defJ}
J_f\left(L^\ast\xi\right)=\langle f, \xi \rangle.
\end{equation}
This functional is bounded on a subspace of $L^2$ by the inequality
\begin{equation}\label{schwartz}
    \left |\langle f, \xi \rangle\right| \leq
    ||f||_{H^{-1}\left(\Omega;x\right)}||\xi||_{H_0^1\left(\Omega;x\right)}
\end{equation}
and by applying Lemma \ref{lemma} to the second term on the right.
Precisely, $J_f$ is a bounded linear functional on the subspace of
$L^2\left(\Omega\right)$ consisting of elements having the form
$L^\ast\xi$ with $\xi\in C_0^\infty\left(\Omega\right).$ Extending
$J_f$ to the closure of this subspace by Hahn-Banach arguments, we
obtain a bounded linear functional defined on all of $L^2.$ The Riesz
Representation Theorem now guarantees the existence of a vector $u\in L^2$ such that
\[
\left(u,L^\ast\xi\right) = J_f\left(L^\ast\xi\right).
\]
But the definition (\ref{defJ}) of the functional $J$ implies that
\[
\left(u,L^\ast\xi\right)=\langle f, \xi \rangle,
\]
which is our definition of a distribution solution.
\end{proof}

In the special case $\kappa = k = 1,$ we can prove the existence of
weak solutions for arbitrary $\mathcal{K}$ satisfying the conditions
of Sec.\ 2.

\begin{theorem} \label{theorem2} The conclusion of Theorem
\ref{theorem1} extends to solutions of the equation
\[
\mathcal{K}(x)u_{xx}+\mathcal{K}'(x)u_x+u_{yy}=f,
\]
where $\mathcal{K}$ satisfies conditions (\ref{cond1}) and (\ref{cond2}).
\end{theorem}

\begin{proof} In the proof of Lemma \ref{lemma}, take
$b_2(y)\equiv 0$ and notice that the term $\beta$ is zero (\emph{c.f.} Proposition \ref{IBP}).
Inequality (\ref{fundalt}) is satisfied for the larger function space
having weight $\mathcal{K}.$ That is, we obtain the inequality
\[
||u||_{H^1_0(\Omega;\mathcal{K})}\leq C||Lu||_{L^2(\Omega)}.
\]
Because $L$ is formally self-adjoint, this is all that is necessary to extend the proof of Theorem \ref{theorem1} to the more general case.
\end{proof}

\appendix

\section{Appendices}

\subsection{A maximum principle for non-uniformly elliptic operators}

The maximum principle referenced in the proof of Theorem \ref{theorem0} is well
known in the case of non-uniformly elliptic operators. For the convenience of the reader
we provide the details, as published proofs tend to assume strict ellipticity. Obviously,
the result is true in greater generality than the form in which we prove it; see the
Remark following Theorem 3.1 of \cite{GT}. The following is a \emph{weak} maximum (minimum) principle, which proves that the supremum  (infimum) of the function occurs on the boundary, but may also occur in the interior. The \emph{strong} maximum (minimum) principle states that if a maximum (minimum) occurs in the interior, the function is a constant. In our case, either form of the maximum principle would give the same result.

\begin{proposition} [essentially due to H. Hopf; see \cite{PW}] \label{max} Let $u\left(x,y\right)$ satisfy the equation
\begin{equation} \label{operdef}
Lu = a\left(x,y\right)u_{xx}+b\left(x,y\right)u_x+ u_{yy}=0,
\end{equation}
where $a\geq 0,$ on a bounded domain $\Omega.$ Then $u$ attains both its supremum and infimum on the boundary $\partial\Omega.$
\end{proposition}

\begin{proof} Suppose that $Lw$ were positive and $w$ attained a maximum at an
interior point $\left(x_0,y_0\right)\in\Omega.$ At that point
we would have
\[
\det\left(
      \begin{array}{cc}
        w_{xx} & w_{xy} \\
        w_{xy} & w_{yy} \\
      \end{array}
    \right)\leq 0
 \]
 and
 \[
 w_x=w_y = 0.
 \]
Because $a\geq0,$ we would also have
\[
\det\left[\left(
            \begin{array}{cc}
              a & 0 \\
              0 & 1 \\
            \end{array}
          \right)
\left(
      \begin{array}{cc}
        w_{xx} & w_{xy} \\
        w_{xy} & w_{yy} \\
      \end{array}
    \right)\right]\leq 0,
 \]
as for any two square matrices $A$ and $B,$
\[
\det AB = \left(\det A\right)\left(\det B\right).
\]
It is also well known (see, \emph{e.g.}, Lemma 8.5 of \cite{Sm}) that if $A$ and $B$ are \emph{symmetric} square matrices with $A\geq 0$ and $B\leq 0,$ then $tr\left(AB\right)\leq 0.$ In our case, this means that
\[
tr\left(
    \begin{array}{cc}
      aw_{xx} & aw_{xy} \\
      w_{xy} & w_{yy} \\
    \end{array}
  \right)=aw_{xx}+w_{yy}\leq 0.
\]
Because $w_x=0$ at $\left(x_0,y_0\right),$ this contradicts our assumption that
\[
Lw = aw_{xx}+bw_x+w_{yy}>0.
\]
We conclude that whenever the operator $L$ of (\ref{operdef}) is strictly positive, then it satisfies a strong maximum principle, and its argument cannot attain a maximum in the interior of its domain.

Let
\[
w = u + \varepsilon e^{\gamma y},
\]
for $\varepsilon$ and $\gamma$ positive. Then
\[
Lw = Lu+\varepsilon L\left(e^{\gamma y}\right) = 0 + \varepsilon \gamma^2 e^{\gamma y} > 0\,\,\forall\varepsilon>0,
\]
so any maximum of $w$ must occur on $\partial\Omega.$ Letting
$\varepsilon$ tend to zero, we conclude that
\[
\sup_\Omega u = \sup_{\partial\Omega} u.
\]
Now at a minimum, $w_{xx}w_{yy}-w_{xy}^2\geq 0.$ So if $Lw<0,$ $w$ cannot attain a minimum at an interior point. We obtain
\[
\inf_\Omega u = \inf_{\partial\Omega} u
\]
by defining
\[
w = u - \varepsilon e^{\gamma y}
\]
and letting $\varepsilon$ tend to zero. This completes the proof.
\end{proof}

\subsection{A weighted Poincar\'e inequality}

The space $L^2\left(\Omega;|\mathcal{K}|\right)$ consists of
functions $u$ for which the norm
\[
||u||_{L^2\left(\Omega;|\mathcal{K}|\right)}=\left(\int\int_\Omega|\mathcal{K}|u^2dxdy\right)^{1/2}
\]
is finite. Standard arguments allow us to define the space
$H^1_0(\Omega; \mathcal{K})$ as the closure of $C_0^\infty(\Omega)$
with respect to the norm
\begin{equation}\label{H101}
    ||u||_{H^1(\Omega; \mathcal{K})}=\left[\int\int_{\Omega}
\left(|\mathcal{K}|u_x^2+u_y^2+u^2\right)\,dxdy\right]^{1/2}.
\end{equation}
The $H^1_0(\Omega; \mathcal{K})$-norm has the form
\begin{equation}\label{H102}
    ||u||_{H^1_0(\Omega; \mathcal{K})}=\left[\int\int_\Omega
\left(|\mathcal{K}|u_x^2+u_y^2\right)\,dxdy\right]^{1/2},
\end{equation}
which can be derived from (\ref{H101}) via a weighted Poincar\'e
inequality. As in the case of Proposition \ref{max}, this inequality is essentially well known (\emph{c.f.} \cite{Sm}, Lemma 4.2 and eq.\ (2.4) of \cite{LMP}), and we include a proof only for the reader's convenience.

\begin{proposition} [Poincar\'e] \label{Poinc} If $u \in H^1_0(\Omega; \mathcal{K}),$ then
\[
||u||^2_{L^2(\Omega)} \leq C\left(\Omega\right)||u||_{H^1_0(\Omega; \mathcal{K})}^2.
\]
\end{proposition}

\begin{proof} It is sufficient to take $u$ to be a continuously differentiable function vanishing on $\partial\Omega,$ and to take $\Omega$ to be the rectangle
\[
    R= \left\{(x,y)|\gamma \leq x \leq \delta, \beta \leq y\leq
\alpha\right\}.
\]
The boundary condition on $u$ allows us to write
\[
u(x,y)=\int_\beta^yu_t(x,t)dt.
\]
By the Schwarz inequality,
\[
u(x,y)
\leq\left(\int_\beta^y\,dt\right)^{1/2}\left(\int_\beta^yu_t^2(x,t)\,dt\right)^{1/2} \leq C \left(\int_\beta^y u^2_t\left(x,t\right)\,dt\right)^{1/2},
\]
where $C$ is a positive constant which depends on $R.$ Squaring both sides (and updating the value of $C$ without changing the notation), we obtain
\[
u^2\left(x,y\right)\leq C\int_\beta^y u^2_t\left(x,t\right)\,dt\leq C\int_\beta^\alpha u^2_t\left(x,t\right)\,dt
= C\int_\beta^\alpha u^2_y\left(x,y\right)\,dy.
\]
If we integrate both sides with respect to $y$ between $\beta$ and $\alpha,$ we multiply the constant $C$ on the right by a new constant which also depends on $R$ (we will also denote the product of these constants by $C$), and obtain on the left the integral of $u$ with respect to $y$ between $\beta$ and $\alpha.$ If we then integrate both sides with respect to $x$ between $\gamma$ and $\delta,$ we obtain
\[
\int\int_R u^2(x,y)\,dxdy \leq C\int\int_R u_y^2(x,y)\,dxdy \leq
C\int\int_R \left(|\mathcal{K}|u_x^2+u_y^2\right)\,dxdy.
\]
This completes the proof of Proposition \ref{Poinc}.
\end{proof}

\subsection{An integration-by-parts formula}

The following result slightly generalizes Proposition 12 of \cite{O1}. It, too, is included for the reader's convenience. The idea of exploiting a formula like the following in order to prove the uniqueness of solutions to (open) elliptic-hyperbolic boundary value problems is apparently due to Friedrichs, but was first applied by Protter \cite{P1}, \cite{P2}; see also \cite{Mo2}. It is known as the \emph{abc-method}. A discussion of this method in the context of equations of Tricomi type can be found in Sec.\ 1 of \cite{Ga}.

\begin{proposition} \label{IBP} Let
\[
Mu = au+bu_x+cu_y,
\]
where $a=$ const., $b = b\left(x,y\right),$ and $c = c(y)$ for smooth functions $b$ and $c.$ Let
\[
L_{\left(\mathcal{K};k\right)}u = \mathcal{K}(x)u_{xx}+k\mathcal{K}'(x)u_x+u_{yy},
\]
where $k$ is a constant. Then the $L^2$-inner product of $Mu$ and $L_{\left(\mathcal{K};k\right)}u$ satisfies
\[
\left(Mu, L_{\left(\mathcal{K};k\right)}u\right) =
\]
\[
\frac{1}{2}\oint_{\partial\Omega} \left(\mathcal{K}(x)
u_x^2+u_y^2\right)\left(cdx-bdy\right)+\int\int_\Omega \omega u^2+\alpha
u_x^2+2\beta u_xu_y+\gamma u_y^2\,dxdy,
\]
where
\[
\omega = \left(1-k\right)\frac{a}{2}\mathcal{K}''(x);
\]
\[
\alpha =
\left[\frac{c_y}{2}-\left(a+\frac{b_x}{2}\right)\right]\mathcal{K}(x)+b\left(k-\frac{1}{2}\right)\mathcal{K}'(x);
\]
\[
2\beta = c\left(k-1\right)\mathcal{K}'(x)-b_y;
\]
\[
\gamma = \frac{1}{2}\left(b_x-c_y\right)-a.
\]
\end{proposition}

\begin{proof}
\[
Mu\cdot Lu =
\left(au+bu_x+cu_y\right)\left(\mathcal{K}(x)u_{xx}+u_{yy}+k\mathcal{K}'(x)u_x\right)
\]
\[
=au\mathcal{K}u_{xx}+auu_{yy}+auk\mathcal{K}'(x)u_x+bu_x\mathcal{K}u_{xx}+bu_xu_{yy}+bu_x^2k\mathcal{K}'(x)
\]
\[
+cu_y\mathcal{K}u_{xx}+cu_yu_{yy}+cu_yk\mathcal{K}'(x)u_x\equiv\sum_{i=1}^9\tau_i.
\]
Taking into account the properties of $a,$ $b,$ $c,$ we have:
\[
\tau_1 = au\mathcal{K}u_{xx} =
\left(au\mathcal{K}u_x\right)_x-au_x^2\mathcal{K}-au\mathcal{K}'(x)u_x=
\]
\[
\left(au\mathcal{K}u_x\right)_x-au_x^2\mathcal{K}-\left(\frac{a}{2}u^2\mathcal{K}'(x)\right)_x+\frac{a}{2}\mathcal{K}''(x)u^2,
\]
using the relation $uu_x=\left(1/2\right)\left(u^2\right)_x;$
\[
\tau_2 = auu_{yy} = \left(auu_y\right)_y-au_y^2;
\]
\[
\tau_3=auk\mathcal{K}'(x)u_x =
\left(\frac{ak}{2}\mathcal{K}'(x)u^2\right)_x-\frac{ak}{2}\mathcal{K}''(x)u^2,
\]
again writing $uu_x$ in terms of the derivative of $u^2;$
\[
\tau_4 =bu_x\mathcal{K}u_{xx}=
b\mathcal{K}\frac{1}{2}\left(u_x^2\right)_x =
\left(\frac{b}{2}\mathcal{K}u_x^2\right)_x-\frac{b_x}{2}\mathcal{K}u_x^2-\frac{b}{2}\mathcal{K}'(x)u_x^2;
\]
\[
\tau_5 =
bu_xu_{yy}=\left(bu_xu_y\right)_y-bu_{xy}u_y-b_yu_xu_y=\left(bu_xu_y\right)_y
\]
\[
-\frac{b}{2}\left(u_y^2\right)_x-b_yu_xu_y=\left(bu_xu_y\right)_y-\left(\frac{b}{2}u_y^2\right)_x+\frac{b_x}{2}u_y^2-b_yu_xu_y;
\]
\[
\tau_6 = k\mathcal{K}'(x)bu_x^2;
\]
\[
\tau_7 =
cu_y\mathcal{K}u_{xx}=\left(cu_y\mathcal{K}u_x\right)_x-cu_y\mathcal{K}'(x)u_x-cu_{yx}\mathcal{K}u_x
\]
\[
=\left(cu_y\mathcal{K}u_x\right)_x-c\mathcal{K}'(x)u_yu_x-\left(\frac{c}{2}\mathcal{K}u_x^2\right)_y+\frac{c_y}{2}\mathcal{K}u_x^2;
\]
\[
\tau_8 =
cu_yu_{yy}=\frac{1}{2}c\left(u_y^2\right)_y=\left(\frac{c}{2}u_y^2\right)_y-\frac{c_y}{2}u_y^2;
\]
\[
\tau_9 = ck\mathcal{K}'(x)u_xu_y;
\]
Integrating over $\Omega$ and collecting terms completes the proof.
\end{proof}

\end{document}